# Sustainable Urban Mobility in the Post-Pandemic Era

Position paper

Christos Theodoridis[1], Yannis Theodoridis[2]

[1] School of Architecture, National Technical University of Athens, Athens, Greece
Email: chteod@mail.ntua.gr

[2] Department of Informatics, University of Piraeus, Piraeus, Greece
Email: ytheod@unipi.gr

## Abstract

COVID-19 is the first pandemic of the modern world causing significant changes to the everyday life of billions of people in all continents. To reduce its expansion, most governments decided to mitigate a great percentage of daily movements of their citizens. For instance, they enforced strict controls (in space, time, etc.) on urban movement whereas they selectively prohibited international air and ground connections. In this short study, we briefly discuss some lessons learned out of this process based on recorded mobility figures, and we raise challenges that are emerging in the post-pandemic era, in the intersection of the sustainable urban mobility and movement data science fields.

## 1. Introduction

Castells (1992) describes how space in todays' era of the 4[th] industrial revolution is defined by flows rather than places, as it used to be considered in its traditional interpretation. Based on this observation, we could consider space as a *place of flows* instead of a *place of places*. Flows could be of any kind: they could be material flows, such as raw materials and commodities, human flows, e.g., people who commute to work or entertainment into a city or even between different continents in a routine basis, as well as immaterial flows, such as information and data. Flows define and preserve the global economy and the living conditions in the modern societies. Flows are of equal importance also at smaller urban scale where the producing traffic mass is enormous. City inhabitants accomplish millions of movements every day with stops at home, workplaces, markets, parks, and entertainment destinations and of course, a high percentage of the total amount of traffic is caused by the necessity of bringing goods to urban centers from the countryside and, vice versa, exporting merchandise from industrial centers to everywhere else.

An important question that arises is what happens when these flows get interrupted, at least those that refer to human movement. If space is defined by flows, then interrupting an important part of them brings as a result a (violent, unprepared) division of space. This is exactly what happened during (and because of) COVID-19, a pandemic that took advantage of the highly developed networks between different regions. In a few months, the virus expanded from China's Wuhan region to all over the world, with rigorousness at the highly networked Southeastern Asia, North America, and Europe. Therefore, COVID-19 is described as a *tertiary domain's virus* (Klaus, 2020). To make a pause at virus' expansion, most countries took measures to mitigate people's movements, at international, national, or even regional scale.

To reduce the expansion of the pandemic, most governments decided to mitigate a great percentage of daily movements of their citizens. Among other measures, they enforced strict controls (in space, time, etc.) on urban movement whereas they prohibited international air and ground connections, on the basis of controlled mobility, according to the "*lower and delay epidemic peak*" model (Bergstrom, 2020).

All means of transport were affected by these restrictions. In the aviation domain, for instance, Heathrow's passengers were reduced by 97% in spring 2020 compared to 2019, making Economist (2020) wonder whether COVID-19 *killed the globalization*; whereas in the maritime domain, the activity decrease in spring 2020 (compared to projections based on past years) was close to –40% for passenger ships (Millefiori *et al*. 2021). Although we admit the significance of other transportation domains, in our study, we focus on the urban mobility and how it was affected by the pandemic.

## 2. Mobility Shift

Although most European countries kept during the pandemics restricting what they considered to be unnecessary movements, e.g., moving from a municipality to another or even moving by car for leisure, some other countries, especially in Eastern Asia, chose a different strategy focusing mainly on individual protection measures and personal diagnostic tests without strict mobility prohibitions.

For instance, Table 1 presents the mobility shift regarding different activities in six example countries, three from Europe (Greece, Italy, and UK) and three from Eastern Asia (South Korea, Singapore, and Taiwan), with mobility controls of very different strictness. The activities include retail & recreation, grocery & pharmacy, parks, transit stations, workplaces, and residential, as recorded during the period Feb. 13, 2020 – Mar. 27, 2021, by Google (2021). European countries, for instance, present dramatic reduction of some activities, even at the level of 50%.

**Table 1: Mobility shift (by activity) in example countries, three from Europe and three from eastern Asia (with and without strict mobility controls, respectively); period: Feb. 13, 2020 – Mar. 27, 2021; baseline refers to Jan. 13, 2020. Data source: Google (2021)**

| activity | GR | IT | UK | ROK | SNG | TW |
|---|---|---|---|---|---|---|
| retail & recreation | –54% | –49% | –55% | –10% | –6% | –7% |
| grocery & pharmacy | +14% | –1% | –5% | +12% | +12% | +3% |
| parks | +45% | –18% | –19% | –13% | –6% | –1% |
| transit stations | –45% | –47% | –54% | –15% | –12% | +1% |
| workplaces | –24% | –27% | –21% | –7% | +4% | –1% |
| residential | +8% | +12% | +11% | +5% | +7% | –1% |

Focusing on three selected countries of Europe (Greece, Italy, and UK), in the paragraphs that follow, we dig into the details of the temporal resolution of mobility changes at urban scale during the above period[1]. Figure 1 illustrates the mobility changes for two capital cities, namely Athens (Greece) and London (UK), selected because of the different approaches the respective authorities followed to face the pandemic, as well as two cities of Italy with very different characteristics, namely Milan and Naples. Three curves are illustrated in the charts, referring to driving (in red), walking (in brown), and transit (in purple), respectively, and this information reflects requests for directions by Apple Maps users (Apple, 2021). Especially for Greece, data for public means of transport are not included due to non-availability.

The charts presented in Figure 1 reveal that during the first wave of pandemic, movement (by any means) degrades dramatically; for instance, in Italy, Apple users' requests were almost zeroed (–90%) for two out of the three transport means, a result of the very strict measures enforced by the Italian authorities. Starting from May, there is a gradual increase of movement, as expected, with a peak during the summer period. On the other hand, during the second wave of pandemic, the lowering pattern appears again, although much smoother than the one that appeared in the first wave. In Mar. 2021, London almost reached the baseline again, while Athens and the two Italian cities still stood below it.

---
[1] The period was selected to cover the two "waves" of pandemic in Europe: the first was roughly from Feb. 2020 to May 2020 and the second was roughly from Oct. 2020 to Mar. 2021. In early Apr. 2021, the percentage of people fully vaccinated in these three countries was about 5%.

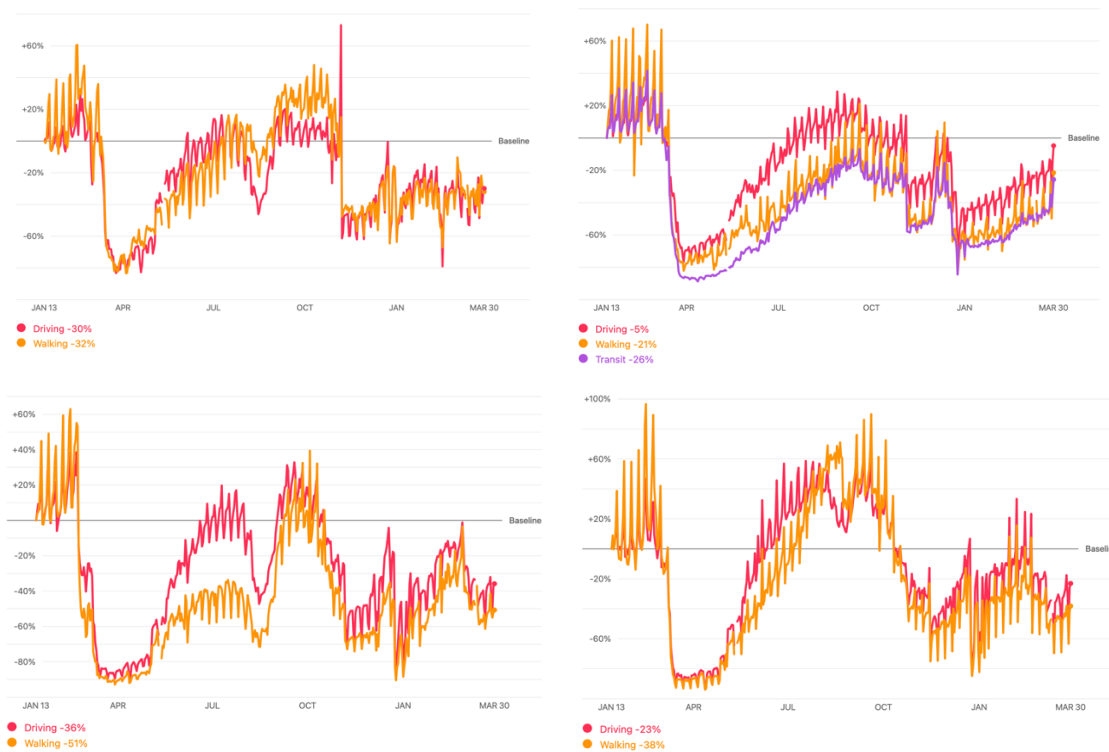

**Figure 1: Mobility variation (by transport means) in four European cities: Athens (top left), London (top right), Milan (bottom left), Naples (bottom right); period: Jan. 13, 2020 – Mar. 27, 2021; baseline refers to Jan. 13, 2020. Data source: (Apple, 2021)**

Considering e.g., Athens as a use case, strict mobility measures were enforced in Mar. 2020, leading to a reduction of –80% for walking and driving. Then, there was the "summer high peak", with the extreme scores of +140% for driving and almost +160% for walking. After-summer period started with an initial slowdown, which was followed by a sharp down due to the second round of strict mobility measures (enforced late Oct. 2020). Since then and until Mar. 2021, driving and walking remained around –40% with respect to the baseline. Furthermore, there are significant differences between cities of the same country. Let us compare, for instance, Milan of the highly industrialized Italian north with Naples of the Italian south: during the period between the two waves, Milan shows a comeback of driving only, whereas in Naples this was the case for walking as well.

Horizontally, the pandemic clearly 'attacked' the public means of transport, and this makes sense because mass transit contradicts social distancing; see e.g., a relevant study in (Rasca *et al.* 2021). Finally, walking is a type of mobility that refers to a place where people are familiar with, like for instance in their neighborhood. Hence, there is a limited need for Google or Apple map apps. So, we can safely infer that walking is even more popular (compared with driving and mass transit) than what is illustrated in the above figures, if we also consider the strict distance restrictions that some governments have applied (during lockdowns in Greece, movement for leisure or retail was constrained inside the same municipality or a 2 km zone).

Overall, there is a correlation between the cities' density of population and connectability with the number of COVID-19 cases, which shows that this *tertiary domain's virus* (Klaus, 2020) is biased towards the big, industrialized, inter-connected urban centers. This is expected to have consequences on the organisation and planning of cities and their urban mobility networks in the post-pandemic era as it seems quite possible that urban life will continue to be affected, at least for the next years, by the pandemic, for instance, due to measures against over-crowded places or because people themselves will try to avoid them.

## 3. New Urban Mobility Approaches

It comes up that the handling of the COVID-19 pandemic is among others an urban planning issue and especially that of transport networks. New mobility schemes emerged at the time of strict mobility prohibitions.

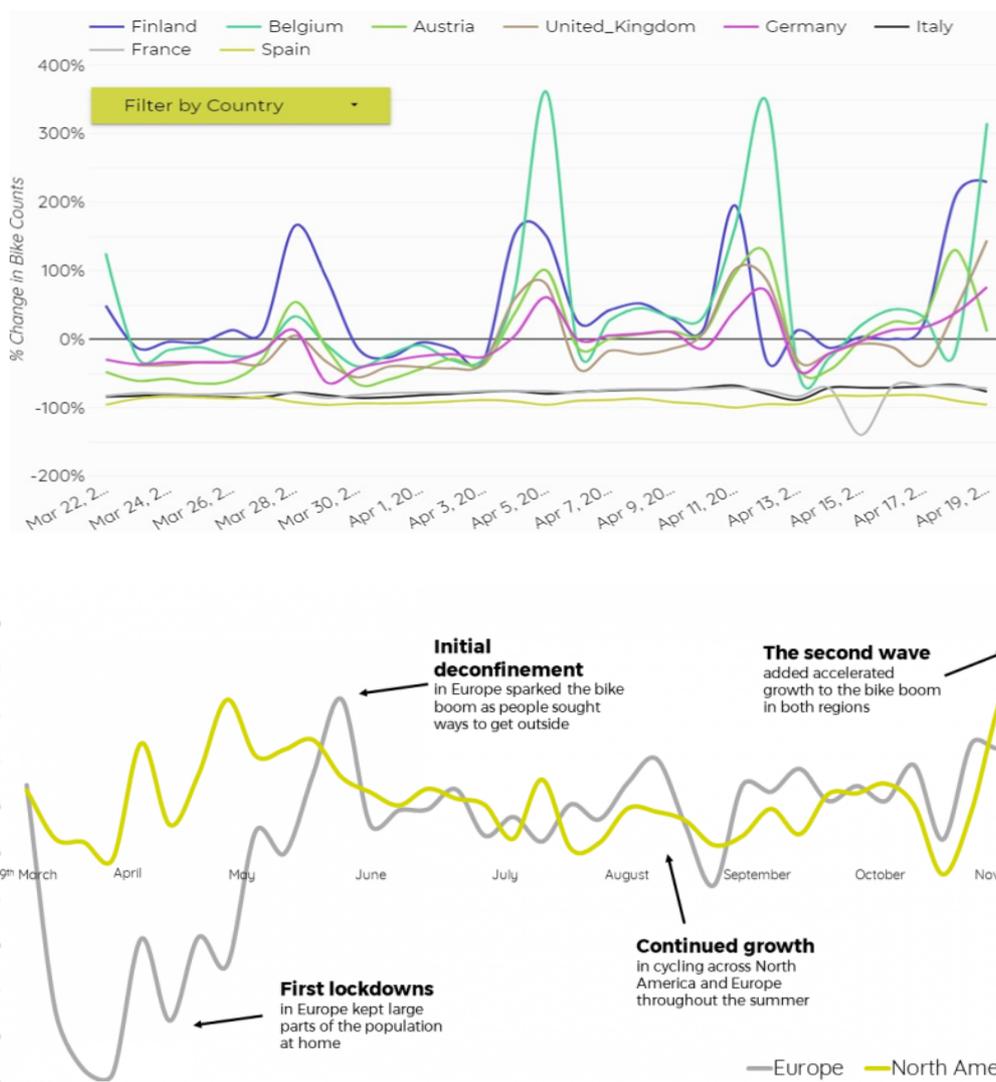

**Figure 2: Use of bicycle variation in eight European countries; period: Mar. 22, 2020 – Apr. 19, 2020 (top); trends in use of bicycle in Europe and North America during the period from the first lockdowns in Mar. 2020 to the second wave in Nov. 2020 (bottom). Source: eco-counter.com**

The chart presented in Figure **2** (top) shows that at the first wave of the pandemic in Europe (spring 2020), in a period of intense mobility prohibitions, in some countries there was a huge increase of bicycle use (+350% in Belgium; +200% in Finland; +100% in UK and Austria), while in others, there was a normal distribution around the baseline and only in Italy, Spain and France there was a significant decrease (-100% in Italy and Spain; -150% in France). Hence, bicycle upgraded its importance and its usage as a means of transport, as illustrated in Figure **2** (bottom) because it achieves social distancing being, at the same time, a healthy habit and an outdoor exercise.

The fact of increased use of bicycle during the pandemic led to an augmentation of awareness about planning and constructing new cycle networks by reducing space from vehicles and deliver it to bicyclists and pedestrians in cities, such as Athens and Rome (Adams, 2020). In Brussels, Milan, Turin, and many UK municipalities, it has been decided to reduce the vehicles' speed limit to 20 or 30 km/h in order to give priority to sustainable means of transport and reduce the risk of traffic accidents (Watson, 2020).

Another aspect that arises is that of *micromobility* and *active mobility* and their relationship with electric means of transport. Active mobility is desired in an era of health instability while micromobility is considered to be an important key at building resilient cities (Naka, 2020). In Paris, the mayor is developing the approach of the *15-minute city*, where every place is accessible with sustainable means in just 15 minutes, which is creating an independence from cars (Sisson, 2020).

## 4. Discussion

During 1980's, some sociologists falsely declared *the end of the cities* because of the new uprising possibilities thanks to digital technology evolution (Parolotto, 2020); until today, people keep accumulating in big urban centers. Nevertheless, during the peaks of the pandemic, movement is more restricted in big urban centers (capitals and big cities) with respect to smaller towns and countryside; this can be justified by the fact that populated places enforce stricter controls and, in cases where horizontal measures are taken at national scale, they are supervised more intensively in places with high density than elsewhere, e.g., in small towns and villages where supervisors and supervisees are very familiar to each other. Moreover, there is a different lifestyle between big cities and small towns or villages: in a large metropolitan area, mobility prohibition drastically changes the daily routine of people during working or leisure time, whereas in places with lower level of urbanization, it is much easier to maintain the same routine by favoring e.g., walking or cycling instead of car driving or using mass transport.

As a consequence, new mobility schemes like micromobility and active mobility have emerged, claiming a key role in this new environment, and some means like bicycle and electronic sustainable vehicles, combined with geospatial information, seem to hold a crucial role in tomorrow's cities' transport networks.

However, a new discussion arises because of COVID-19 that questions the last decades established values in urban planning. For example, the ideal of compact city with high density of land-uses and population may not seem as appealing as it used to be. At the same time, urban villages can be easier isolated and protected than an expanded metropolis. A proposal that may overcome these aspects is that of (ISOCARP, 2020) that refers to sub-regions that are highly inter-connected with public means of transport and at the same time quite resilient and self-reliant so that their inhabitants are not depended on movements to other regions.

An important factor for the proper function of all the above is to combine them with geospatial data science. That could give the solution that is needed to make e.g., public means of transport appealing again, something that got lost during the pandemic. GIS mapping and data availability (and fusion) can be applied to achieve a better supervise of the real-time use and availability of cars, wagons, and buses (Acuto, 2020). Thus, commuters could make everyday decision making on their mobility according to their need of speed, time, and safety, based on geospatial data with high precision.

Especially in what concerns sustainable urban mobility, in the post-pandemic era it is very probable to face a new and very different reality from what we used to know. Under these circumstances, COVID-19 effects raise challenges in the post-pandemic era, in the intersection of sustainable urban mobility (Tumlin 2012; Johung and Sen 2013; Lah 2019) and geospatial, in particular movement data science (Pelekis and Theodoridis 2014; Dodge *et al*. 2021; Andrienko *et al*. 2021).

Concluding, mobility effects due to pandemic are of core attention for urban planners and geospatial data scientists because they can define not only economic ratios but also the everyday life of people all over the world and especially at its most urbanized part. In a pandemic, *mobility is at the same time the victimizer and the victim*. Even if the global economy is restored after pandemics, a return to a previous situation regarding mobility sounds quite uncertain.